\documentclass[twocolumn,showpacs,amsmath,amstex,amssymb,citeautoscript,longbibliography]{revtex4-1}
\pdfoutput=1
\usepackage[english]{babel}
\usepackage{letltxmacro}

\LetLtxMacro{\ORIGselectlanguage}{\selectlanguage}
\makeatletter
\DeclareRobustCommand{\selectlanguage}[1]{%
  \@ifundefined{alias@\string#1}
    {\ORIGselectlanguage{#1}}
    {\begingroup\edef\x{\endgroup
       \noexpand\ORIGselectlanguage{\@nameuse{alias@#1}}}\x}%
}
\newcommand{\definelanguagealias}[2]{%
  \@namedef{alias@#1}{#2}%
}
\makeatother
\definelanguagealias{en}{english}
\definelanguagealias{English}{english}
\usepackage{graphicx}
\usepackage{amsmath}
\usepackage{amsfonts}
\usepackage{amssymb}
\usepackage{bm}
\usepackage{color}

\usepackage{soul} 
\usepackage{amssymb}
\usepackage{wasysym}
\usepackage{dsfont}
\usepackage{multirow}
\usepackage{epsfig}

\usepackage{hyperref}
\hypersetup{
    bookmarks=false,         
    unicode=false,          
    pdftoolbar=false,        
    pdfmenubar=true,        
    pdffitwindow=false,     
    pdfstartview={FitH},    
    pdftitle={A criterion for many-body localization-delocalization phase transition},    
    pdfauthor={Maksym Serbyn, Z. Papic, and Dmitry A. Abanin},     
    pdfsubject={},   
    pdfcreator={},   
    pdfproducer={}, 
    pdfkeywords={many-body localization} {matrix elements} {disordered systems}, 
    pdfnewwindow=true,      
    colorlinks=true,       
    linkcolor=black,          
    citecolor=blue,        
    filecolor=magenta,      
    urlcolor=blue           
}

\newcommand{\be}{\begin{equation}}
\newcommand{\ee}{\end{equation}}
\newcommand{\bea}{\begin{eqnarray}}
\newcommand{\eea}{\end{eqnarray}}

\newcommand{\la}{\langle}
\newcommand{\ra}{\rangle}

\renewcommand{\phi}{\varphi}
\renewcommand{\epsilon}{\varepsilon}

\newcommand{\corr}[1]{{\langle #1\rangle}}

\newcommand{\1}{\mathds{1}}

\usepackage{color}

\begin{document}
\title{A criterion for many-body localization-delocalization phase transition}

\author{Maksym Serbyn$^1$, Z. Papi\'c$^2$, and Dmitry A. Abanin$^{3,4}$}
\affiliation{$^1$ Department of Physics, University of California, Berkeley, California 94720, USA}
\affiliation{$^2$ School of Physics and Astronomy, University of Leeds, Leeds, LS2 9JT, United Kingdom}
\affiliation{$^3$ Department of Theoretical Physics, University of Geneva, 24 quai Ernest-Ansermet, 1211 Geneva, 
Switzerland}
\affiliation{$^4$ Perimeter Institute for Theoretical Physics, Waterloo, ON N2L 2Y5, Canada}
\date{\today}

\begin{abstract}

We propose a new approach to probing ergodicity and its breakdown in quantum many-body systems based on their response to a local perturbation. We study the distribution of matrix elements of a local operator between the system's eigenstates, finding a qualitatively different behaviour in the many-body localized (MBL) and ergodic phases. To characterize how strongly a local perturbation modifies the eigenstates, we introduce the parameter ${\cal G}(L)=\la \ln ({V_{nm}}/{\delta}) \ra$, which represents a disorder-averaged ratio of a typical matrix element of a local operator $V$ to the energy level spacing, $\delta$; this parameter is reminiscent of the Thouless conductance in the single-particle localization. We show that the parameter ${\cal G}(L)$ decreases with system size $L$ in the MBL phase, and grows in the ergodic phase. We surmise that the delocalization transition occurs when ${\cal G}(L)$ is independent of system size, ${\cal G}(L)={\cal G}_c\sim 1$. We illustrate our approach by studying the many-body localization transition and resolving the many-body mobility edge in a disordered 1D XXZ spin-$1/2$ chain using exact diagonalization and time-evolving block decimation methods. Our criterion for the MBL transition gives insights into microscopic details of transition. Its direct physical consequences, in particular logarithmically slow transport at the transition, and extensive entanglement entropy of the eigenstates, are consistent with recent renormalization group predictions.  

\end{abstract}


\maketitle

\section{Introduction}\label{Sec:intro}

 Experimental advances of the past decade have lead to the realization of isolated quantum many-body systems of cold atoms and trapped ions. These systems have slow intrinsic time scales and are amenable to various quantum-optics experimental probes, thus they are ideally suited for probing far-from-equilibrium quantum dynamics~\cite{PolkovnikovRMP}. The studies of cold atoms and trapped ions have inspired a theoretical quest for the universal framework to describe non-equilibrium quantum phenomena. One of the central challenges that has emerged is to understand the microscopic mechanisms of ergodicity and its breakdown in isolated quantum systems~\cite{PolkovnikovRMP}.

It has been established that there exist two distinct generic classes of many-body systems: the {\it ergodic} ones, which thermalize as a result of quantum evolution, and the {\it many-body localized} (MBL)~\cite{Basko06,Mirlin05,OganesyanHuse,PalHuse}, which are disordered and avoid thermalization via a mechanism similar to Anderson localization~\cite{Anderson58} in the Hilbert space. Ergodic and many-body localized systems have drastically different spectral and dynamical properties. Microscopic mechanism of thermalization in ergodic systems is called the Eigenstate Thermalization Hypothesis (ETH)~\cite{SrednickiETH,RigolNature}. According to the ETH, thermalization in ergodic systems is manifest already in the properties of individual eigenstates, in which physical observables are thermal. In contrast, the eigenstates of MBL systems violate the ETH due to the emergence of extensively many local integrals of motion (LIOMs)~\cite{Serbyn13-1,Huse13,Imbrie14,ScardicchioLIOM,Chandran14}. 

One of the invaluable tools for probing ergodicity that has recently emerged is the entanglement entropy. In ergodic systems, the eigenstates are similar to random vectors in the Hilbert space (modulo global conservation laws, e.g. energy); therefore, the eigenstates have an extensive, volume-law entanglement entropy. In contrast, the eigenstates of MBL systems can be obtained from product states by quasi-local unitary transformations, and have low entanglement that obeys the ``area law''~\cite{Serbyn13-1,Bauer13}. Further, ergodic and MBL phases are distinguished by the dynamics of entanglement  following a quantum quench from initially non-entangled states: in the former case, entanglement spreads ballistically and grows linearly in time~\cite{LiebRobinson,Kim13}, while in the latter case the spreading is logarithmic in time~\cite{Moore12,Serbyn13-2,Kim14}. The latter property has been understood from the phenomenological theory of the MBL phase based on the LIOMs~\cite{Serbyn13-1,Huse13}.

Recently, first signatures of MBL have been observed in systems of cold atoms~\cite{Bloch15}. In this experiment, disorder can be tuned in a broad range, which allows one to probe MBL and ergodic phase, as well as the transition between them. The MBL-delocalization transition is a new kind of a dynamical phase transition, across which the eigenstates undergo a dramatic transformation as their entanglement entropy changes from area-law to volume-law. Although it is conceivable that certain many-body systems exhibit an intermediate non-ergodic phase, here we focus on the case of a direct transition between the MBL and ergodic phases, which is suggested by numerical studies of 1D systems~\cite{PalHuse,Serbyn13-2,Luca13,Kjall14,Rigol14,Alet14,Sirker14,Reichman14,Scardicchio15}, as well as by a recent phenomenological renormalization group study~\cite{AltmanRG14}, and an effective percolation model~\cite{Potter15}. Developing a full theory of this transition remains a major challenge. 

In this paper, we propose a new approach to studying MBL, ergodicity, and the transition between them. We introduce a single dimensionless parameter defined in terms of physically measurable quantities -- matrix elements of local operators -- which provides a direct and sensitive probe of MBL. We formulate a criterion of MBL-delocalization transition in terms of this parameter, and explore the physical properties at the critical point. The parameter can be viewed as a many-body generalization of the Thouless conductance~\cite{Thouless72} in the single-particle Anderson problem; the latter plays a central role in the scaling theory of localization~\cite{ScalingTheory}.

\begin{figure}[t]
\begin{center}
\includegraphics[width=0.999\columnwidth]{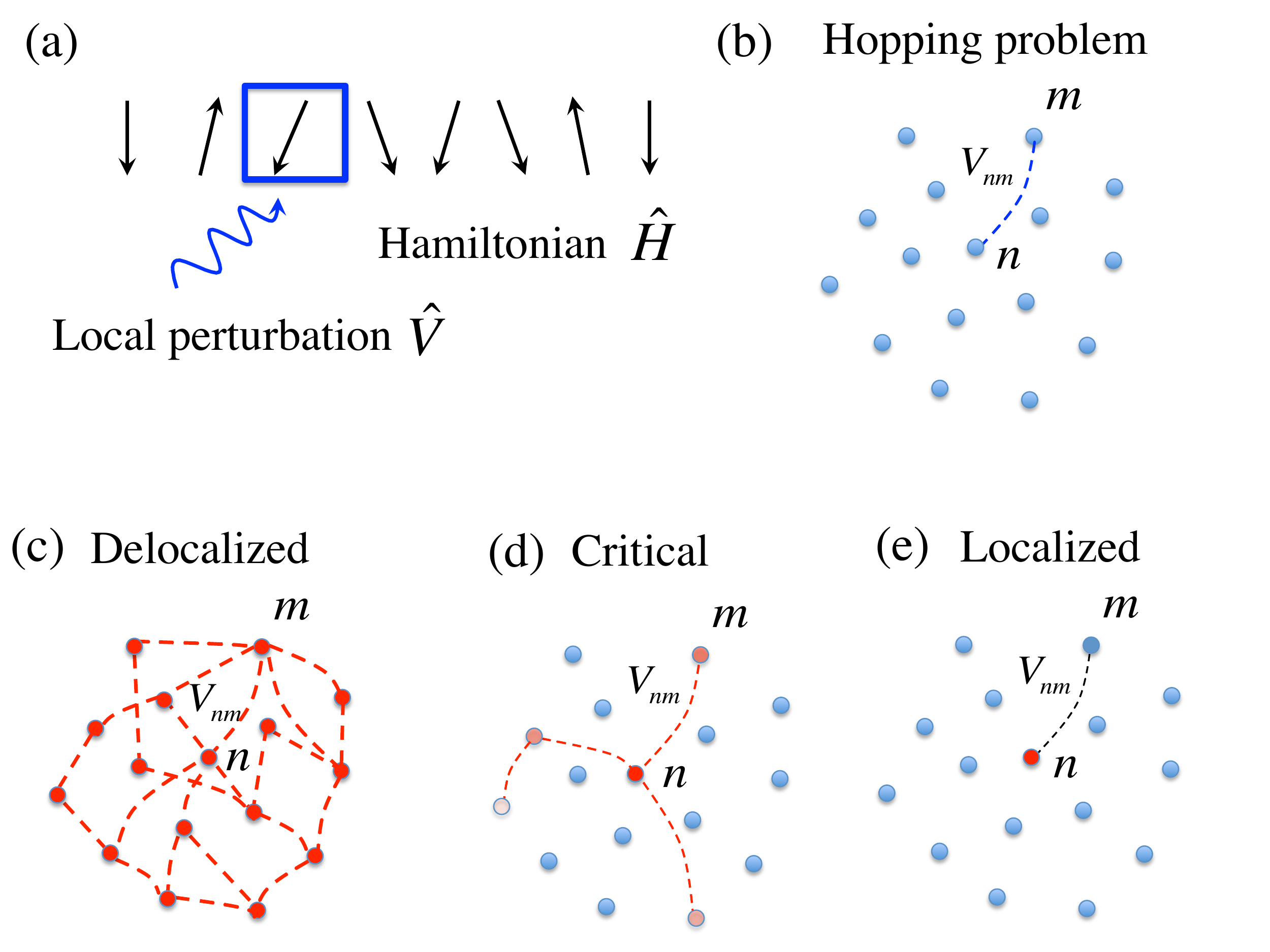}\\
\caption{ \label{Fig:cartoon}  (a) The system is perturbed by a local physical operator $\hat V$ that mixes the eigenstates of the unperturbed Hamiltonian $\hat H$, leading to  an effective hopping problem shown in the panel (b). Vertices of the lattice correspond to the unperturbed eigenstates, and matrix elements of the operator $\hat V$ determine the hopping amplitudes.  Bottom panels illustrate the possible outcomes of a local perturbation applied to the Hamiltonian: all the eigenstates are mixed in the ergodic phase (panel c), whereas in the MBL phase the perturbation admixes a finite number of states (panel e).  In the critical regime, the eigenstates are expected to be multifractal~\cite{Gruzberg2008}.}
\end{center}
\end{figure}

The main idea of our approach, summarized in Fig.~\ref{Fig:cartoon}, is to study the response of the system's eigenstates to a local perturbation $V$.  We consider a lattice many-body system with an unperturbed Hamiltonian $H$ and eigenstates $|n\ra$, $H|n\ra=E_{n}|n\ra$. The problem of finding the spectrum and eigenstates of $H+V$ can be viewed as a hopping problem on a lattice where sites correspond to the eigenstates $|n\ra$ of unperturbed Hamiltonian $H$, with on-site energies $E_n'=E_n+V_{nn}$ and hopping amplitudes $V_{nm}$ between sites $n,m$ (see Fig.~\ref{Fig:cartoon}). Thus, all the information is encoded in eigenenergies $E_n'$ and in the off-diagonal matrix elements $V_{nm}$. Note that for local $V$, $V_{nn}\lesssim ||V||$, and in general the order of energies $E_{n}'$ is different from $E_{n}$. 

A key insight is that perturbing the Hamiltonian with a {\it local} operator $V$ modifies the eigenstates very differently in the MBL and ergodic phases~[see Fig.~\ref{Fig:cartoon}(a-b)]. In the MBL phase, the perturbation $V$ can only affect the degrees of freedom within a localization length $~\xi$ away from its support. Thus, the perturbation strongly admixes a given eigenstate with only a finite number of other eigenstates [Fig.~\ref{Fig:cartoon}(e)], as expected from the existence of LIOMs in the MBL phase. In contrast, in the ergodic phase $V$ hybridizes an extensive number of eigenstates, therefore its effect is highly non-local, and the eigenstates of $H+V$ are completely delocalized in the basis of the unperturbed eigenstates. 

To detect the localization/delocalization transition, we ask whether hopping induced by $V$ hybridizes the eigenstates with neighbouring values of $E_n'$ (the closest states in the many-body spectrum). To characterize this hybridization, we introduce the parameter
\be\label{eq:g}
\mathcal{G}(\epsilon,L)=\ln \frac{|V_{n,n+1}|}{E_{n+1}'-E_{n}'}, 
\ee
where we have ordered the eigenstates by their modified energy $E_n'$, and $\epsilon=E_n'/L= E_n/L+\mathcal{O}(1/L)$ is the energy density.  As suggested above, the scaling of $\mathcal{G}$ with system size $L$ should be qualitatively different in the MBL and ergodic phases. 

In the MBL phase, at $L\gg \xi$ the eigenstates close in energy typically have very different spatial structure due to their different values of LIOMs, and a local operator couples them exponentially weakly in $L$. Therefore, in the MBL phase $\mathcal{G}(L)\propto -L$. On the other hand, in the ergodic phase the ETH implies that a local perturbation mixes the energy levels very strongly, and this requires $\mathcal{G}(L)\propto +L$. The above considerations suggest a natural criterion for the delocalization transition at the energy density $\epsilon$ to be $\la \mathcal{G}(\epsilon,L) \ra \sim 1$. This condition implies that a local perturbation significantly changes the structure of the eigenstates, such that the LIOMs become non-local and the effective hopping problem enters the critical regime. 
These results will be analytically and numerically substantiated in Sections~\ref{Sec:Analytic} and \ref{Sec:Numerics} below. It will be shown that ${\mathcal{G}}$, as a single parameter, characterizes the delocalization transition.

Previous numerical studies of the delocalization transition have relied on the statistics of energy levels~\cite{PalHuse}, as well as the fluctuations of entanglement entropy~\cite{Kjall14}. In contrast, our approach allows us to directly probe the localization properties of the eigenstates through the matrix elements of local operators. Unlike entanglement entropy and level statistics, these are directly related to correlation functions, and therefore are, in principle, measurable. Moreover, our approach yields further consequences for the properties of the system at the critical point. Those include the extensive entanglement in the eigenstates, and logarithmic in time transport of conserved quantities and entanglement propagation. These properties are qualitatively consistent with the predictions of recent phenomenological renormalization group studies~\cite{AltmanRG14,Potter15}. 

The rest of the paper is organized as follows. In Section~\ref{Sec:Analytic} we provide analytical derivation for the scaling of $\cal G$ with energy density and system size in the MBL and ergodic phases, and formulate the delocalization criterion. Using exact diagonalization, in  Section~\ref{Sec:Numerics} we then test our approach numerically on the random-field XXZ spin chain, where the transition between MBL and ergodic phases can be driven by changing the disorder strength. We demonstrate that the $\mathcal{G}(L)\approx 1$ criterion gives a location of the transition that is in good agreement with the estimates obtained using other methods~\cite{Alet14}. Furthermore, we show that our approach has a good energy-density resolution, and apply it to map out the phase diagram as a function of disorder and energy density, demonstrating the presence of the many-body mobility edge in the random-field XXZ model. In  Section~\ref{Sec:consequences} we address the physical consequences for the MBL transition that follow from our analysis. In particular, we find the logarithmic in time propagation of entanglement and particle number fluctuations, in agreement with recent works~\cite{AltmanRG14,Potter15}. 
We numerically verify those predictions using exact diagonalization and time-evolving block-decimation simulations of a global quench in the random-field XXZ model. Our conclusions and a discussion of future directions is presented in Section~\ref{Sec:Summary}.

\section{Statistics of matrix elements and the scaling of $\mathcal{G}$: analytic considerations}\label{Sec:Analytic} 

In this Section we discuss the distribution of the parameter $\mathcal{G}$ introduced in Eq.~(\ref{eq:g}) and its scaling in the MBL and ergodic phases. In the former case, the behaviour of $\mathcal{G}$ can be understood by invoking the ETH. In the latter, our considerations are based on the picture of LIOMs in MBL phase. In the subsequent Section, we present data from exact diagonalization which are fully consistent with these analytical arguments. 

\subsection{Ergodic phase: matrix elements from the ETH}

In the ergodic phase, the eigenstates are expected to be similar to random 
vectors in the Hilbert space (this analogy is qualitative due to locality of interactions and energy conservation).

Consider two eigenstates $|n\ra, |m\ra$ near the middle of the many-body band. They satisfy the orthogonality constraint $\la m|n\ra=0$, $n\neq m$. Intuitively, acting on the state $|n\ra$ with a local perturbation, $V|n\ra$, gives another random vector in the Hilbert space, which is no longer orthogonal to $|m\ra$. Thus, it is expected that the off-diagonal matrix element $V_{nm}=\la m|V|n\ra$ is a scalar product of two random vectors, $|m\ra$ and $V|n\ra$. Denoting the Hilbert space dimension by $\mathcal{D}$, the matrix element can be estimated as $V_{nm}\sim 1/\sqrt{\mathcal D}$. For a quantum many-body system, in general  $\mathcal{D}\sim \exp( s L)$, where $L$ is the number of lattice sites or spins, and $s$ is statistical entropy density.  Since the level spacing scales as the inverse Hilbert space dimension, $\delta \sim 1/{\mathcal D}$, we have $\delta \ll V_{nm}$, and a local perturbation strongly mixes the nearby eigenstates. Our parameter $\mathcal{G}$, which in the leading order is given by $\mathcal{G}(L)\sim \ln ({V_{nm}}/{\delta}) \sim {1}/{2}\ln {\mathcal D}\propto  L$,  thus grows linearly with system size.

The intuitive argument above can be extended to states with an arbitrary energy density using Srednicki's~\cite{Srednicki99}  ansatz for the off-diagonal matrix elements of local operators: 
\be\label{eq:srednicki_ansatz}
V_{nm}=e^{-S(E,L)/2} f(E_n,E_m) R_{nm}, 
\ee
where $S(E)$ is the statistical entropy at energy $E=(E_n+E_m)/2$, $R_{nm}$ is a random number of order one, and $f$ is a smooth function which can be linked to the thermally-averaged response of the system to the perturbation $V$. 
Since $\delta\propto e^{-S(E)}$, we obtain that
\be\label{eq:matrix_ergodic}
\mathcal{G}(L)=\ln\frac{V_{n,m}}{\delta_{n}} \propto \frac{S(E,L)}{2} \propto +\frac{s(\epsilon)}{2} L, 
\ee
 for states with an arbitrary energy density $\epsilon = E/L$, where $s(\epsilon)$ is the corresponding entropy density. It is worth noting that the off-diagonal matrix elements in the ergodic phase have been studied numerically in Ref.~\cite{Haque15}, where the scaling $V_{nm}\propto 1/\sqrt{\mathcal{D}}$ was confirmed.

\subsection{MBL phase: matrix elements from local integrals of motion \label{Sec:VMBL}}

To understand  the properties of the off-diagonal matrix elements $V_{nm}$ and $\mathcal{G}$ deep in the MBL phase we rely on the picture of LIOMs~\cite{Huse13,Serbyn13-1,Imbrie14,ScardicchioLIOM}. It was argued that the key property of a system that shows many-body localization at infinite temperature is that its eigenstates can be connected to product states by a quasi-local unitary transformation $U$, which entangles the remote degrees of freedom only exponentially weakly. 

For concreteness, let us consider an interacting spin-1/2 model [e.g., the XXZ model introduced in Eq.~(\ref{eq:hamiltonian}) below]. In this case, it is convenient to choose the natural ``up-down'' basis, i.e., the basis vectors are eigenstates of all $\sigma_i^z$ operators. Then, in the MBL phase there exists a quasi-local unitary operator $U$ (quasi-local means that this operator only creates short-range entanglement), which diagonalizes the Hamiltonian in this basis, $U^\dagger HU=H_{\rm diag}$. 
The unitary $U$ defines a mapping from the physical spin operators, $\sigma_i^\alpha$, to ``effective spins'' $\tau_i^\alpha$ via
\be\label{eq:tauz}
\tau_i^\alpha=U\sigma_i^\alpha U^\dagger.
\ee
The operators $\tau^z_i$ form a complete set of mutually commuting, quasi-local integrals of motion in the MBL phase. In the $\tau$-basis the eigenstates therefore simply become ``up-down'' states of the effective spins.

In order to analyze the effect of a local perturbation, we expand the corresponding local operator $V$ (which for simplicity is chosen to act on the first few spins in the chain)  over the complete basis of operators $\tau_{1}^{\alpha_1} \tau_{2}^{\alpha_2}...\tau_{L}^{\alpha_L}$, where $\alpha_i \in \{0,\pm,z\}$. Here $\tau_i^0=\1$ denotes the identity operator on the site $i$, and $\pm$ denote the usual raising/lowering operators. We represent $V$ as a sum over operators $V_r$ of range $r$, which affect only the effective spins $1,\ldots,r$:  
\be\label{eq:Vexpansion}
V=\sum _r V_r, \quad V_r=\sum_{\alpha_1,\ldots,\alpha_r} V_r^{\alpha_1\ldots\alpha_r} \tau_1^{\alpha_1}\ldots\tau_r^{\alpha_r}, 
\ee
where the last $\alpha_r\neq 0$. The off-diagonal matrix element of $V$ between the eigenstates which differ by $\sim L$ flips of  effective spins is equal to $V_L^{\alpha_1...\alpha_L}$, where $\sim L$ coefficients are $\alpha_i=\pm$ such that the operator $\tau_1^{\alpha_1}\ldots\tau_r^{\alpha_r}$ maps one configuration into another. For example, if $|n_1\rangle = \uparrow \downarrow \uparrow \downarrow$, and $|n_2\rangle = \downarrow \uparrow \downarrow \uparrow$, the matrix element $\langle n_2|V|n_1\rangle = V^{-+-+}$ gives us a coefficient in front of $\tau_1^-\tau_2^+\tau_3^-\tau_4^+$ term in the expansion~(\ref{eq:Vexpansion}).  Thus the matrix elements of $V$ probe the nature of the spatial decay of the coefficients $V_L^{\alpha_1...\alpha_L}$, which, in turn, encode the properties of the unitary $U$ (in particular, they control the amount of entanglement generated by $U$). 

Deep in the MBL phase, the LIOMs are  robust under local perturbations. Typically, $V$ hybridizes a given eigenstate only with states which have the same integrals of motion distance $x\lesssim \xi$ ($\xi$ being the localization length) away from where $V$ acts. Two generic eigenstates typically have different integrals of motion everywhere. Hence, in order for a local perturbation $V$ to couple these states, it has to flip the state of effective spins across the entire volume of the system. This requires tunneling an ``excitation'' through the entire system, which would leave behind a trace of spin flips~\footnote{This is essentially the picture of a quasiparticle moving and creating particle-hole excitations which was proposed in  Refs.~\cite{Levitov97,Mirlin05}.}. In the localized phase, such tunneling is exponentially suppressed; as a consequence, the probability of $V$ to couple  such states decays exponentially with system size. Thus, typically the ratio $V_{nm}/\delta$ behaves as $e^{-\kappa L}$, and the parameter $\cal G$ decays linearly with the system size:
\be\label{eq:g_mbl}
{\cal G}(L)\propto -\kappa L. 
\ee
Below we study ${\cal G}(L)$ numerically for the XXZ spin chain, finding the behaviour consistent with the above formula. We also find that ${\cal G}(L)$ has a broad, nearly normal distribution, similar to the distribution of the wave function tails in the single-particle Anderson insulator~\cite{MirlinRMP}.

Note, that Eq.~(\ref{eq:g_mbl}) indicates that matrix elements of a local perturbation are similar to a random ultrametric matrix~\cite{Fyodorov2009}. Indeed,  the scaling of $\cal G$ in Eq.~(\ref{eq:g_mbl}) implies that a local perturbation which changes values of the LIOM up to a distance $x$ away from its center~[there are $2^{x-1}$ such eigenstates], has a matrix element $V(x)\propto e^{-(\kappa+s)x}$, where $s$ is the entropy density. For $\kappa>0$ the eigenstates are localized, and the transition occurs at $\kappa=0$~\cite{Fyodorov2009}, which is similar to our criterion. The problem of Anderson localization for random ultrametric matrices has been studied analytically using strong-disorder renormalization group~\cite{Levitov90}, and, in particular, multifractal exponents have been computed~\cite{Fyodorov2009}. Note, however, the im-
portant difference of the ensemble of ultrametric matrices
considered in the literature, is that the matrix elements there
obey the gaussian, rather than the log-normal distribution. We reserve the exploration of this analogy to further studies. 

We note that the behavior of matrix elements in the MBL phase was used~\cite{Rahul14} to study the properties of a local spectral function. In addition, the spatial decay of the {\it diagonal} part of local operators given by terms with all $\alpha_i \in  \{0,z\}$ in Eq.~(\ref{eq:Vexpansion}) was studied in Ref.~\cite{Chandran14}, and was also found to decay exponentially in the MBL phase.

\subsection{A criterion for the transition}

Using the predicted linear scaling of the parameter $\mathcal{G}$ with system size in the ergodic and MBL phases, it is natural to propose a phenomenological criterion for the delocalization transition as the absence of scaling of $\mathcal{G}$ with $L$:
\begin{equation} \label{Eq:criteria}
 \mathcal{G}(L) = \mathcal{G}_c.
\end{equation}
Below, we will use this criterion to determine the location of MBL-ergodic transition as a function of disorder strength and energy density in exact diagonalization. 

Physically, the above criterion implies that LIOMs, characteristic of the MBL phase, become unstable and are expected to delocalize at the transition. Indeed, any local perturbation strongly couples a given eigenstate with an adjacent  eigenstate (in the reshuffled spectrum $E_n'$), which has a different spatial structure, and therefore different LIOMs. In this case, the new eigenstates are superpositions of states with different values of the LIOMs across the entire system. Thus, a local perturbation modifies the integrals of motion even far away from where it is applied, which implies that integrals of motion necessarily become non-local. 

We have argued that the above condition implies the non-locality of integrals of motion.  It is natural to ask whether the criterion~(\ref{Eq:criteria}) indeed implies delocalization in the effective hopping problem we introduced in Section~\ref{Sec:intro} above, and what are the properties of the system at this delocalization transition. That is, whether the participation ratio of new eigenstates in the basis of unperturbed eigenstates tends to infinity as $L\to \infty$. One cannot answer this without additional information regarding the distribution of matrix elements. We have studied several quantities which probe the delocalization in the hopping problem directly, and indeed found that the above criterion is consistent with the delocalization in the hopping problem.  

\begin{figure*}[t]
\begin{center}
\setlength{\unitlength}{\columnwidth}
\begin{picture}(0,0)
\put(-0.01, 0.5){\large (a)}
\put(0.25, 0.48){$W=0.5$}
\put(0.65, 0.5){\large (b)}
\put(0.9, 0.48){$W=3.6$}
\put(1.3, 0.5){\large (c)}
\put(1.55, 0.48){$W=5$}
\put(-0.02, 0.22){\rotatebox{90}{$p({\cal G})$}}
\put(0.65, 0.22){\rotatebox{90}{$p({\cal G})$}}
\put(1.3, 0.22){\rotatebox{90}{$p({\cal G})$}}
\put(0.33, -0.02){$\cal G$}
\put(.95, -0.02){$\cal G$}
\put(1.63, -0.02){$\cal G$}
\end{picture}
\includegraphics[width=0.6\columnwidth]{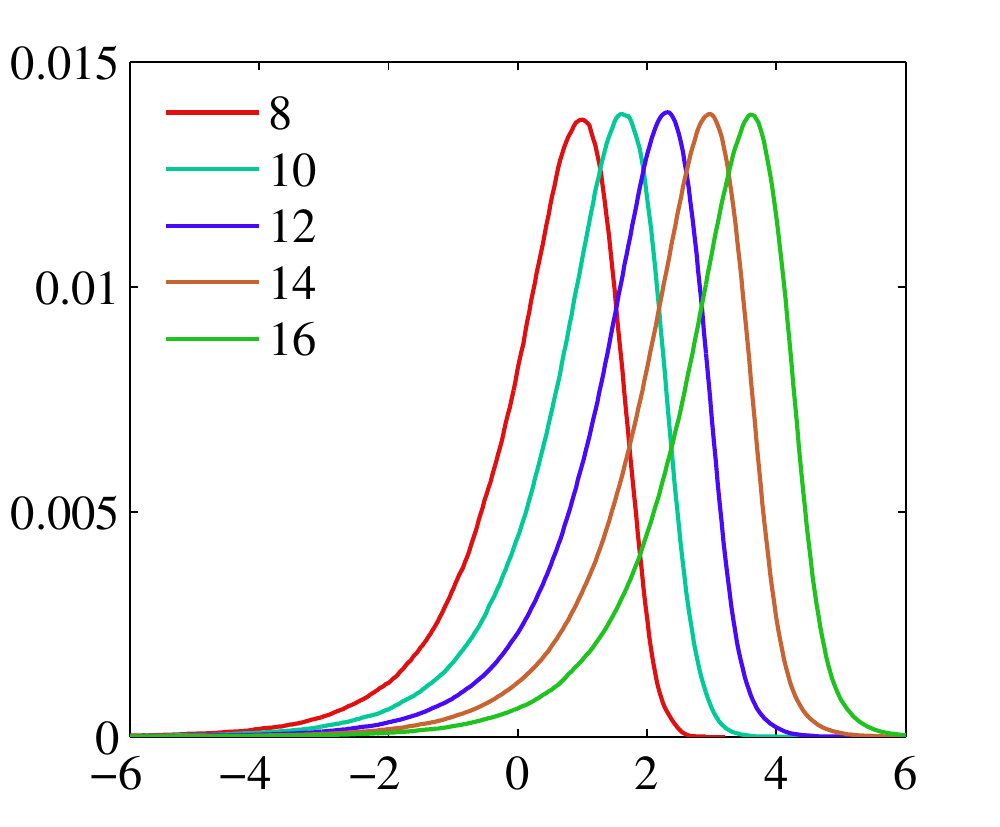}
\hspace{5pt}
\includegraphics[width=0.6\columnwidth]{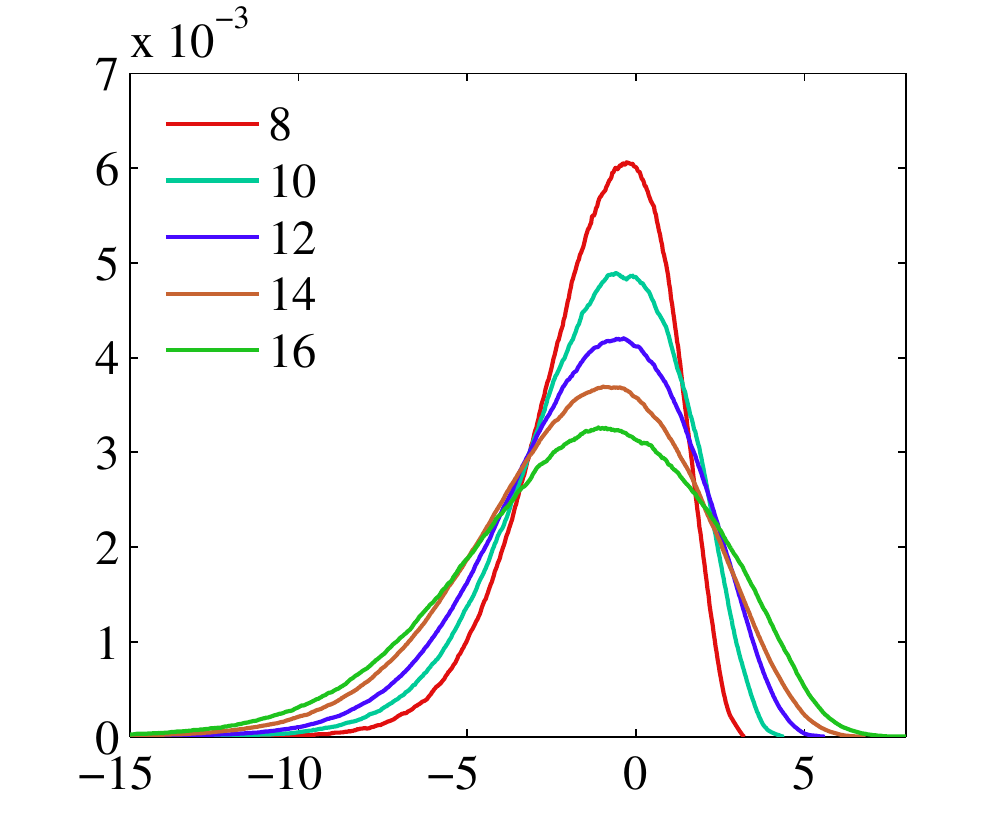}
\hspace{5pt}
\includegraphics[width=0.6\columnwidth]{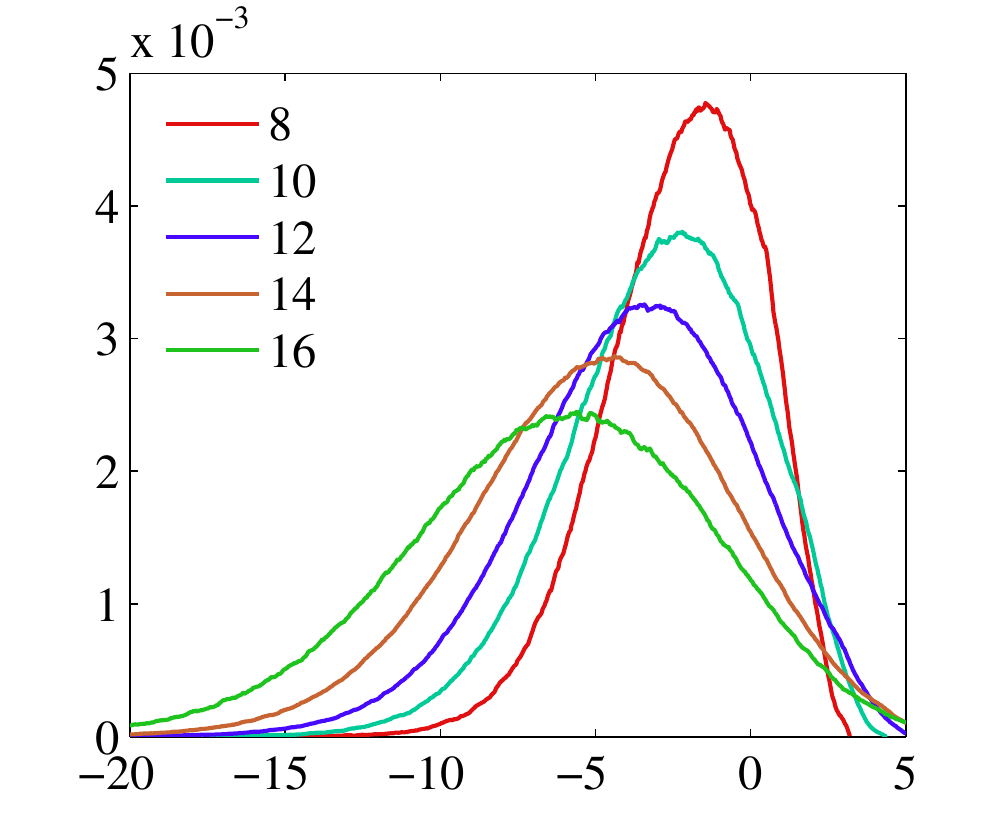}
\caption{ \label{Fig1} 
 Distribution of ${\cal G}= \log \left(|V_{n n+1}|/\delta\right)$  across the MBL transition displays qualitatively different scaling with system size.  (a) For weak disorder ($W=0.5$),when the system is in the ergodic phase, $\cal G$ increases with system size, and the distribution shifts to the right.  (b) At the MBL transition ($W= 3.6$), the distribution broadens but does not move. (c) In the MBL phase ($W=5$),  $\cal G$ becomes smaller for larger systems, and the shape of the distribution is approximately gaussian.}
\end{center}
\end{figure*}
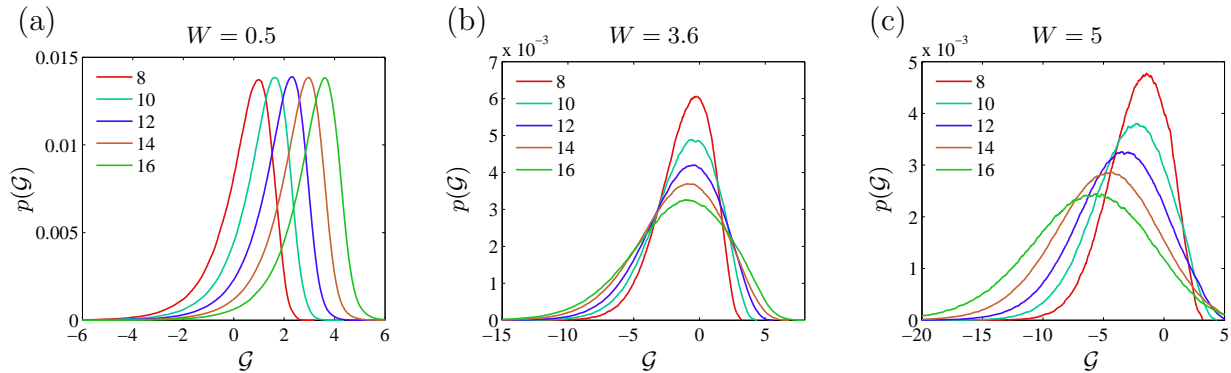 
\section{Numerical results \label{Sec:Numerics}}

In this Section we test the predictions for the distribution and scaling of $\mathcal{G}$ using numerical (exact) diagonalization of the one-dimensional random-field XXZ spin-$1/2$ chain. The Hamiltonian of this model is given by:
\be\label{eq:hamiltonian}
H= J_x\sum _{i=1}^{L}  (\sigma_i^+ \sigma_{i+1}^- +\sigma_i^- \sigma_{i+1}^+)+
      \frac12 J_z\sum _{i=1}^{L} \sigma_i^z \sigma_{i+1}^z +
      \sum_i^{L} h_i \sigma_i^z, 
\ee
where we use periodic boundary conditions and identify $\sigma_{L+1}\equiv \sigma_1$. Below we set $J_x=1$. The random $z$-field $h_i$ on each site is drawn from a box distribution $[-W;W]$, and $\sigma^\pm = (\sigma^x\pm i\sigma^y)/2$. At $J_z=0$, this model can be mapped onto free spinless fermions moving in a disorder potential via the Jordan-Wigner transformation; in this case, all eigenstates are localized for arbitrary disorder strength $W$. Introducing $J_z\neq 0$ is equivalent to turning on the nearest-neighbour interactions between fermions. 

Previous work~\cite{PalHuse} has demonstrated that the model~(\ref{eq:hamiltonian}) exhibits the MBL phase at strong disorder ($W>W_c$), and the ergodic phase at weak disorder ($W<W_c$). The delocalization transition takes place at some critical disorder strength $W_c$. Several numerical studies~\cite{PalHuse,Luca13,Rigol14,Sirker14,Alet14,Reichman14,Bardarson2015,Scardicchio15,Chamon15} have estimated the location of the transition in this model using various probes, including the statistics of eigenenergies, entanglement entropy and its fluctuations, participation ratios and different dynamical probes. In particular, state-of-the-art exact diagonalization performed on systems up to $L=22$ spins~\cite{Alet14} has robustly determined the location of the transition at $W_c\approx 3.6$. 

Here we use exact diagonalization to numerically study the statistics of the off-diagonal matrix elements and parameter ${\mathcal{G}}$ for the XXZ spin chain (\ref{eq:hamiltonian}) across the MBL-delocalization transition. In order to establish the universality of this approach, we considered the response of the spin chain to two different local perturbations:
\be\label{eq:V}
\hat V_1=\sigma_I^z, 
\quad 
\hat V_2=\sigma_I^+ \sigma_{I+1}^-+\sigma_I^- \sigma_{I+1}^+. 
\ee
The interaction strength was fixed to be $J_z = 1$, similar to previous studies. All the data were obtained for chains with periodic boundary conditions constraining the Hilbert space to states with zero total spin, such that the dimension is ${\mathcal D}(L) = {L\choose{L/2}}$.  Disorder averaging was performed over at least $10^4$, $5000$, $4000$, $2000$ and $200$ realizations for $L=8,10,12,14,16$ spins.

\subsection{Distribution of ${\cal G}(L)$ in the middle of the band}

First we study the {\it disorder-averaged} distribution of $\mathcal{G}$ over ${\mathcal D}(L)/L$ eigenstates in the middle of the band. As discussed in the introduction, we perturb the eigenspectrum by the diagonal matrix elements of the chosen operator, $E_n' = E_n+\corr{n|\hat V_{1,2}|n}$. Perturbed eigenenergies $\{E_n'\}$ are sorted, and we collect the statistics for $\mathcal{G}(L)$ as defined in Eq.~(\ref{eq:g}).

The behaviour of the distribution function of $\mathcal{G}(L)$ is illustrated in Fig.~\ref{Fig1}. In this case, the perturbation operator was chosen to be $\hat V_1$ (we found a qualitatively similar behaviour for $\hat{V}_2$). 
We observe that the distribution of $\mathcal{G}$ exhibits a qualitative change across the MBL transition. In the ergodic phase~[Fig.~\ref{Fig1}(a)], the average value of $\la \mathcal{G}(L) \ra$ grows with the system size, signalling that a local perturbation is more and more likely to hybridize the nearby states -- a signature of delocalization. If one scales the distribution by the square root of the Hilbert size dimension,  $\la \mathcal{G}(L) \ra /\sqrt{\mathcal D}$, the distributions for different system sizes indeed approximately collapse, confirming that $\mathcal{G}$ scales as predicted by the ETH.

In contrast, in the MBL phase~[Fig.~\ref{Fig1}(c)], the averaged value of $\la \mathcal{G} \ra$ decreases as system size is increased; this reflects the fact that a local perturbation is less and less likely to hybridize the nearby states, and therefore the eigenstates and the LIOMs remain robust. Moreover, the distribution of $\mathcal{G}$ becomes broader at larger $L$, and is approximately normal. We found that the distribution becomes very close to normal at large system sizes and strong disorder. 

Finally, at intermediate disorder $W=3.6$, the average $\la \mathcal{G}(L) \ra$ remains nearly independent of system size. Hence it is natural to identify this point with the MBL-delocalization transition.  Below we will use the energy-resolved average value of $\mathcal{G}$ to identify the transition point and the mobility edge. Note that although the average value of $\la \mathcal{G}(L) \ra$ remains constant at the transition, the distribution broadens as the system size is increased. Detailed investigations revealed that this effect originates largely from sample-to-sample fluctuations, rather than from the state-to-state fluctuations in the same disorder realization. 

\subsection{Average energy-resolved $\corr{{\cal G}(\varepsilon,L)}$}

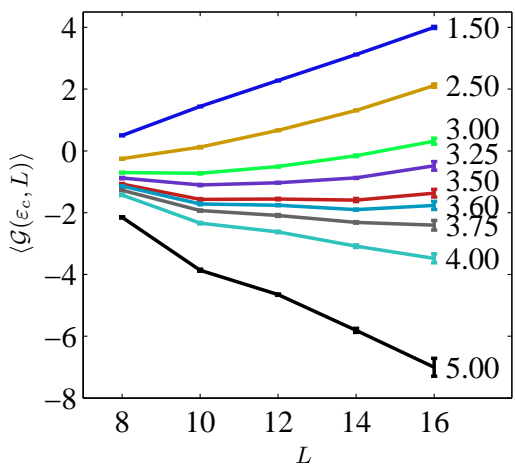
\begin{figure}[b]
\begin{center}
\setlength{\unitlength}{\columnwidth}
\begin{picture}(0,0)
\put(0.01, 0.3){\rotatebox{90}{$\corr{{\cal G}(\epsilon_c,L)}$}}
\put(0.45, -0.02){$L$}
\end{picture}
\includegraphics[width=0.85\columnwidth]{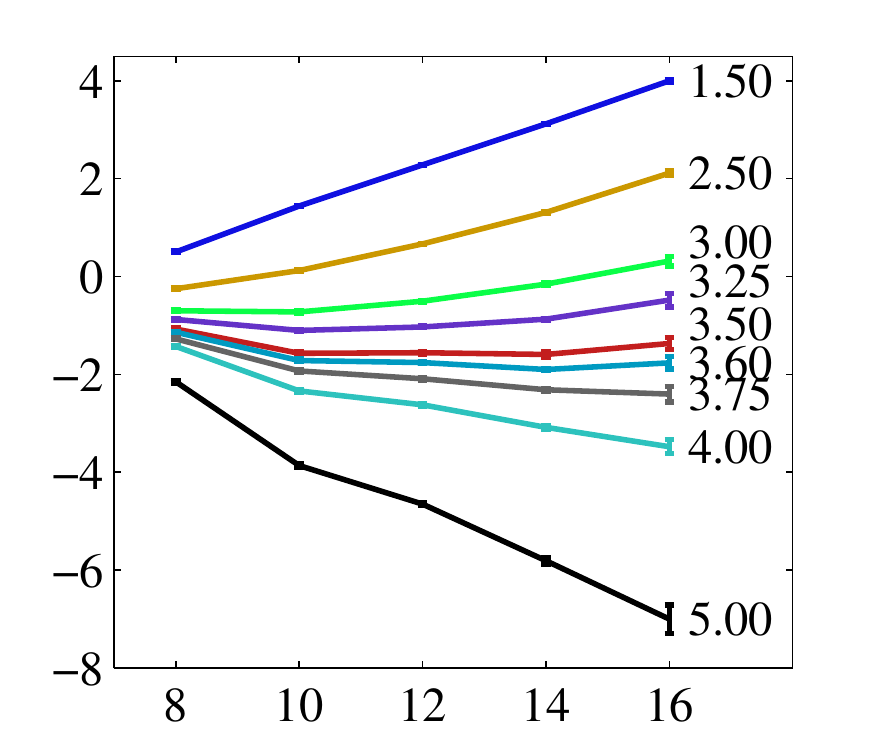}
\caption{ \label{Fig2} 
Scaling of $\mathcal{G}(\varepsilon_c,L)$ in the middle of the band with the system size for different disorders $W=1.5,\ldots, 5$. Value of disorder is shown on the right of each curve. From here the critical disorder strength is determined as $W_c=3.6\pm 0.15$.}
\end{center}
\end{figure} 

In order to obtain the location of the MBL transition more precisely, we now study the average value $\corr{\mathcal{G}(\varepsilon,L)}$ as a function of the dimensionless energy density and system size. The dimensionless energy density $\varepsilon $ (referred to as ``energy density'' in the following, for simplicity) is defined as $\varepsilon = (E-E_\text{min})/(E_\text{max}-E_\text{min})$, where $E_\text{min/max}$ are the energies of ground state (highest excited state) of our system.  

Figure~\ref{Fig2} shows the system-size dependence of $\corr{{\cal G}(\epsilon_c,L)}$ for fixed $\epsilon_c=0.45$, which is the energy density at which the delocalized phase is most robust. Similar to the behaviour already observed for the distribution of $\cal G$, we see that the behaviour of averaged $\corr{{\cal G}(\epsilon_c,L)}$ is qualitatively different at weak and strong disorder. At $W\lesssim 3$ we have  ${d\corr{{\cal G}(\epsilon_c,L)}}/{dL}>0$, and the second derivative appears to be positive, signalling that larger systems become more and more thermal. At strong disorder $W\geq 4$, 
$\corr{{\cal G}(\epsilon_c,L)}$ behaves according to Eq.(\ref{eq:g_mbl}), as expected in the MBL phase. From  Fig.~\ref{Fig2} we identify the critical value of disorder $$W_c =3.6\pm0.15$$ using our criterion for the MBL transition, Eq.~(\ref{Eq:criteria}). This agrees with the previous findings of Refs.~\cite{Alet14}. 

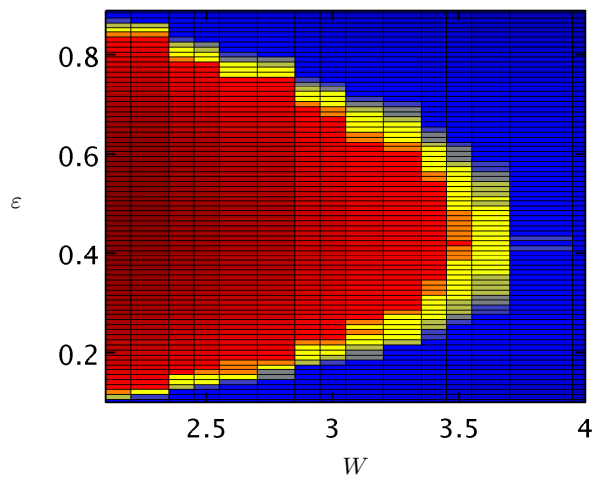
\begin{figure}[b]
\setlength{\unitlength}{\columnwidth}
\begin{center}
\begin{picture}(0,0)
\put(0.5, -0.03){$\Large W$}
\put(-0.01, 0.38){$\Large \varepsilon$}
\end{picture}
\includegraphics[width=0.95\columnwidth]{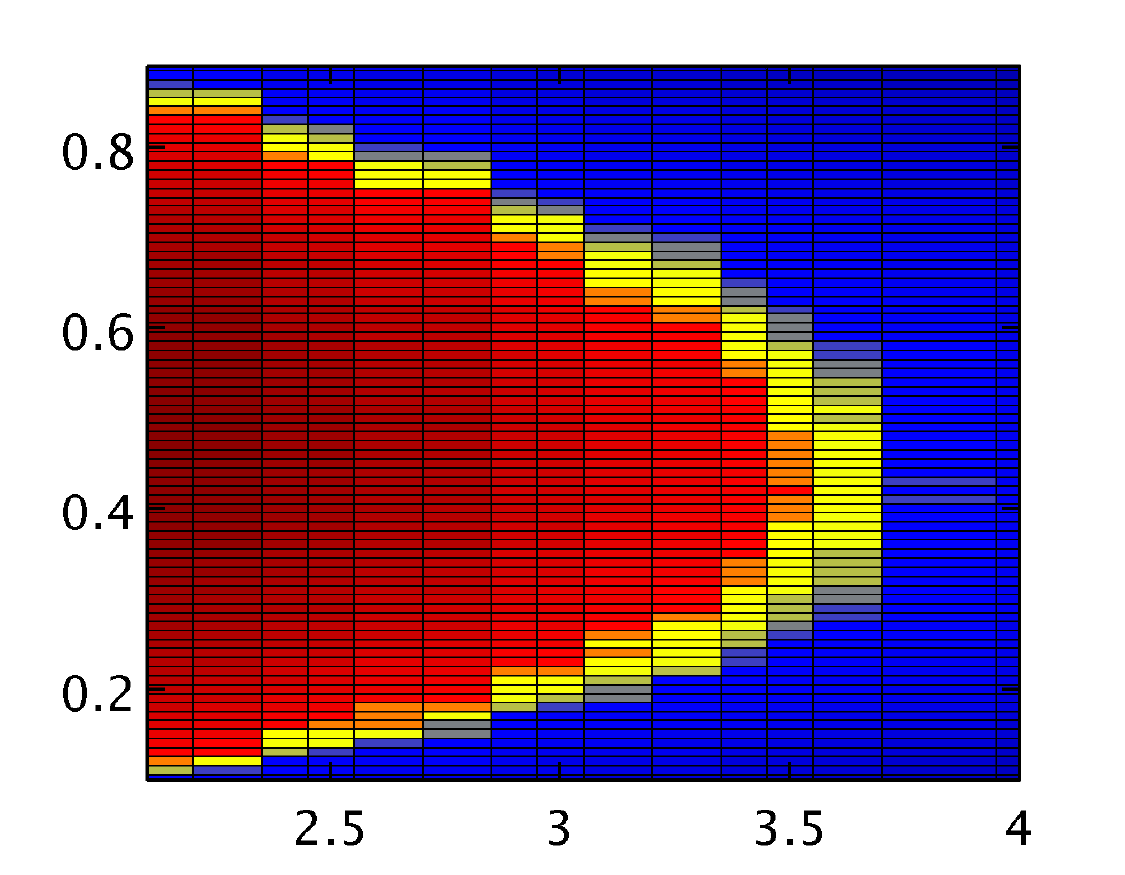}
\caption{ \label{Fig3}  Many-body mobility edge $\varepsilon(W)$ as a function of disorder. Blue (red) color indicates regions where $\corr{g(\epsilon,L)}$ decreases (grows) with $L$. Yellow regions correspond to points where we cannot determine the behavior due to errorbars.  
}
\end{center}
\end{figure} 

Using the same delocalization criteria, we can map out the many-body mobility edge in the random-field XXZ model. We define the many-body mobility edge to be at the energy density $\varepsilon(W)$ where $\corr{{\cal G}(\epsilon_c,L)}$ is independent of system size.  Results of this calculation are shown in Fig.~\ref{Fig3}.  We also notice the intrinsic asymmetry of the mobility edge: it is shifted towards the ground state and centered slightly below $\epsilon=1/2$~\cite{Alet14,Hughes15,Bardarson2015}. This asymmetry becomes stronger with increasing the interaction strength (not shown), which agrees with the recent arguments of Ref.~\cite{Hughes15}. 

\section{Entanglement and dynamics at the delocalization transition \label{Sec:consequences}} 

Next, we explore the physical properties of the system at the transition, defined using the above criterion. In particular, we will be interested in understanding the entanglement properties of the eigenstates, as well as the dynamical properties -- spreading of entanglement and transport of conserved quantities. We discuss qualitative expectations and support them with numerical studies. 

 In order to qualitatively understand the entanglement properties of the eigenstates, we consider a system of size $2L$ and initially disconnect it across the middle link into left and right subsystems. The eigenstates of such a system are simply product states of the eigenstates in the left and right parts, $|n\ra_L \otimes |m\ra_R$, $n,m=1,...,{\mathcal D}$. Restoring the coupling between $L$ and $R$ systems corresponds to a local perturbation $V_{LR}=\sum_\alpha V_{\alpha L}\otimes V_{\alpha R}$, where the sum includes a small finite number of terms (three for the case of the XXZ spin chain), and $V_{\alpha L}, V_{\alpha R}$  act only on the degrees of freedom in the left/right part. The problem of finding the eigenstates reduces to a hopping problem, where the sites are product states $|n\ra_L \otimes |m\ra_R$, and hopping amplitudes are set by the operator $V_{LR}$. This problem bears many similarities with the case of a local perturbation acting on a single system, discussed above. 

Previously in Section~\ref{Sec:Analytic} we established that the MBL transition corresponds to the critical regime of the effective hopping problem. In this case, we expect that the eigenstates of the connected system $|I\ra$ are multifractal when expressed in the product basis $|I\ra=\sum_{n,m} A_{nm}^I |n\ra_L \otimes |m\ra_R$, with disorder-averaged participation ratios satisfying a relation $P_q=\sum_{n,m} |A_{nm}^I|^{2q}\propto {\mathcal D}^{-2\tau_q}$. This is consistent with recent work~\cite{Santos15} which studied dynamical signatures of the fractality near the MBL transition.
We have tested and confirmed the multifractality of the hopping problem numerically (details will be presented elsewhere). Although we are not aware of a direct relation between entanglement entropy and participation ratios in the many-body case (for the single-particle problem such a relation does exist, see Ref.~\cite{Gruzberg2008}), it is natural to expect that the entanglement entropy of half the system will be extensive, but sub-thermal:
$$
S_\text{ent}(L)\sim \alpha s L, 
\quad  \alpha<1,
$$
where $sL$ is the maximal (thermal entropy) of the subsystem of size $L$. This agrees with the predictions of Ref.~\cite{AltmanRG14}. We note that the above relation is expected to hold for entanglement between two regions of approximately equal size. If, on the other hand, one of the subsystems is much smaller than the other one, its entanglement entropy should approach the thermal value, in accordance with the general arguments based on the strong subadditivity of entanglement~\cite{Grover14}.

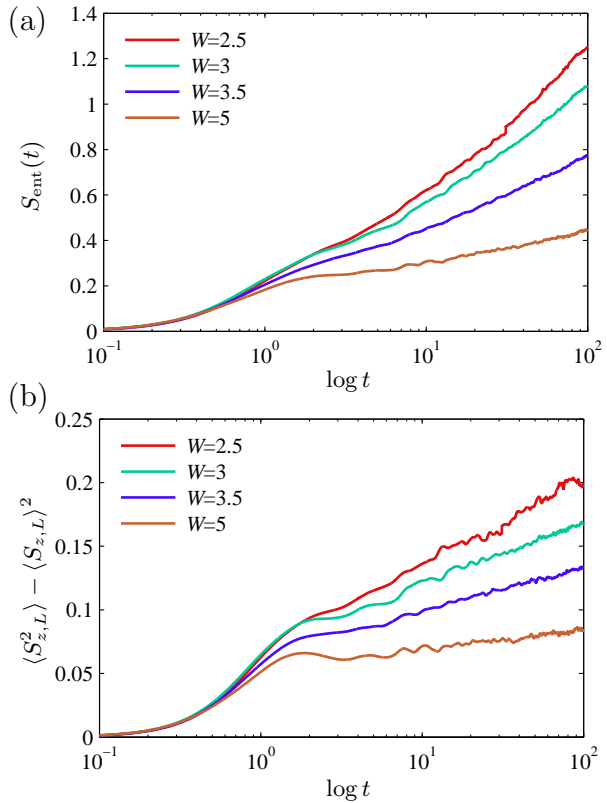
\begin{figure}[b]
\setlength{\unitlength}{\columnwidth}
\begin{picture}(0,0)
\put(0.48, -0.02){$\Large \log t$}
\put(0.48, -0.65){$\Large \log t$}
\put(-0.01, 0.53){\large (a)}
\put(-0.01, -0.05){\large (b)}
\put(0.01, 0.25){\rotatebox{90}{$S_\text{ent}(t)$}}
\put(0.01, -0.45){\rotatebox{90}{$\langle S_{z,L}^2 \rangle - \langle S_{z,L} \rangle^2$}}
\end{picture}
\includegraphics[width=0.96\columnwidth]{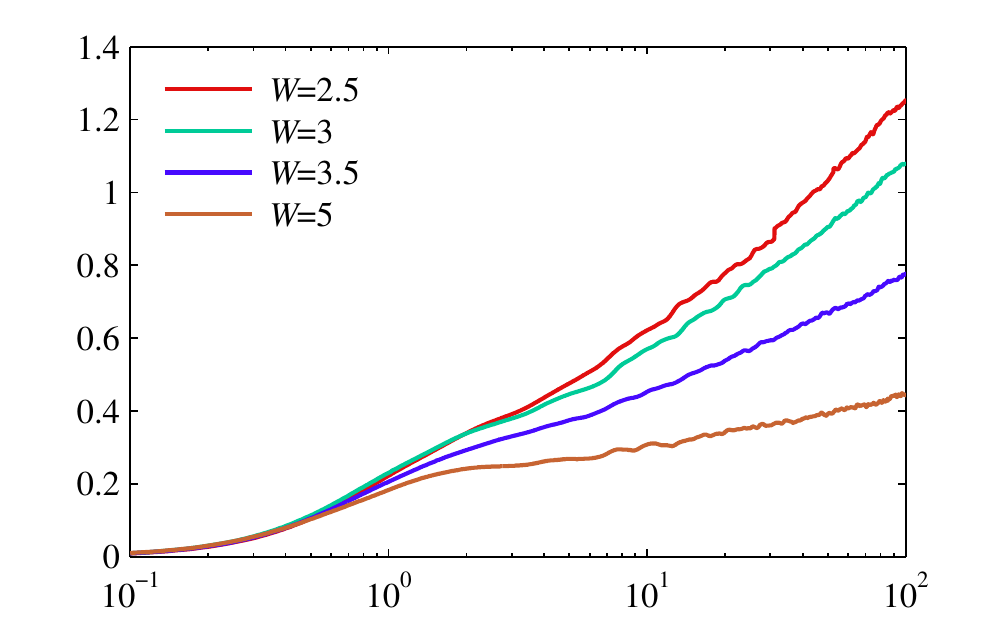}\\
\vspace{5pt}
\includegraphics[width=0.96\columnwidth]{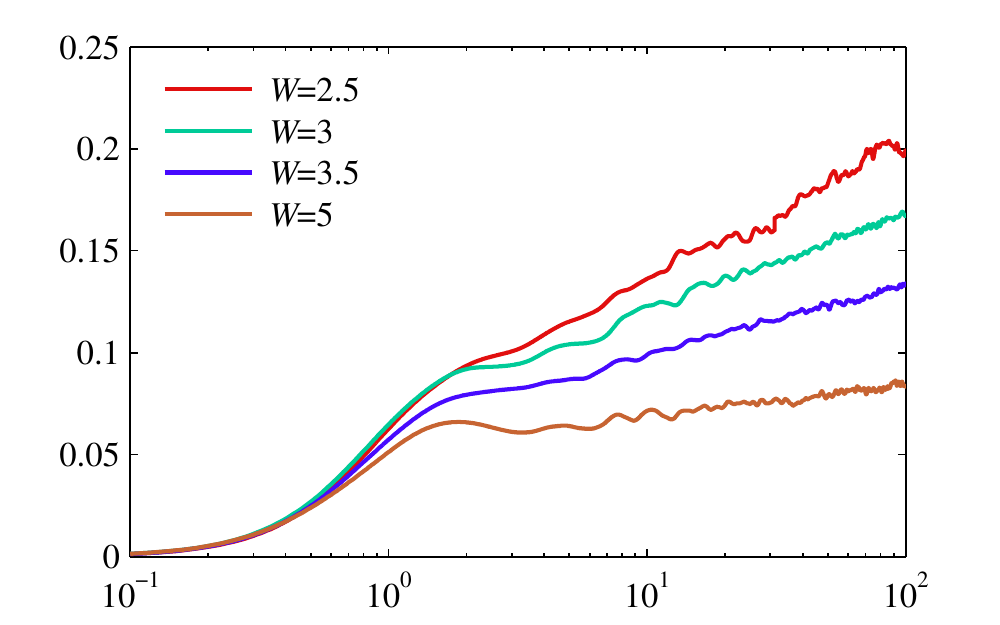}
\caption{ \label{Fig4} Dynamics of entanglement entropy (a) and particle number fluctuations (b) across the MBL transition. The system was initialized in a product state where each spins is up or down. For entanglement entropy, the transition corresponds to faster-than logarithmic growth of entanglement. The particle number fluctuations grow logarithmically in the vicinity of the transition.}
\end{figure}  

Next, we discuss the dynamics at the transition. Assuming the analogy with the problem of ultrametric random matrices, described above (although it should be kept in mind that this analogy is almost certainly incomplete -- in particular, in our case the off-diagonal matrix elements follow a log-normal distribution rather than a random distribution), we expect that at the transition the typical relaxation time scale for a system of size $L$ is exponentially long in $L$,   
\begin{equation}
\tau(L)\sim \frac{1}{\delta(L)} = e^{s L},
\end{equation}
and therefore the transport at the transition will be logarithmically slow,  consistent with the  predictions of Refs.~\cite{AltmanRG14,Potter15}. Note that this is a time scale for both \emph{dephasing}, and the \emph{on-shell dynamics} (transport) at the MBL transition. For example, let us consider a setup in which the left and right subsystems are initially disconnected, i.e., are prepared in an eigenstate $|\psi(0)\ra=|n_0\ra_L\otimes |m_0\ra_R$, and the coupling between them is turned on instantaneously at $t=0$. In the MBL phase after a quench, due to dephasing, the entanglement entropy increases logarithmically, and this growth is unbounded in an infinite system; but there is no transport (except for rare resonances). At the transition, however, there is transport of the $z-$component of spin polarization, which  can be characterized by its fluctuations in the left subsystem as a function of time. Hence, both fluctuations of the magnetization and entanglement growth are unbounded in the thermodynamic limit and increase logarithmically in time,
\be\label{eq:dynamics}
S_{\rm ent}(t)\sim \ln t,
\qquad 
 \langle S_{z,L}^2 \rangle - \langle S_{z,L} \rangle^2 \sim \ln t, 
\ee
where $S_{z,L}$ is the total $z-$projection of spin in the left part. We expect that these laws will be generic and will hold for a class of initial states, including the case when the system is prepared in a product state (rather than in a product state of eigenstates of the left and right parts). 

We have studied the entanglement growth and spin transport numerically to verify  Eq.~(\ref{eq:dynamics}). We considered a global quench in which the system is initially prepared in a product state (random up-down configuration of spins), and time evolved with the Hamiltonian in Eq.~(\ref{eq:hamiltonian}). Using exact diagonalization, we have been able to access long-time dynamics in systems of up to $L=18$ spins. In addition, we have performed time-evolving block decimation~\cite{tebd} (TEBD) simulations which allow access to much larger spin chains, but are restricted to moderate times $t\sim 10^2$ \footnote{Calculations were performed using the ITensor Library, http://itensor.org/.}.  For the TEBD algorithm we used the Trotter decomposition with time step $\Delta t=0.01$ and a maximum bond dimension $\chi=2000$. The results on the entanglement growth and the fluctuations of the spin polarization, obtained using TEBD for $L=24$ spins, are illustrated in Fig.~\ref{Fig4}. Both quantities spread logarithmically in time. We note that slow transport opens an additional channel for entanglement growth, in addition to dephasing~\cite{Serbyn13-2}, and as a result, the entanglement $S_\text{ent}(t)$ near the transition does not display a characteristic ``knee'' which corresponds to the crossover between transport-dominated and dephasing-dominated growth deeper in the MBL phase.  

\section{Summary \label{Sec:Summary}}

We presented a new approach for studying the MBL and delocalization transition based on the response of system's eigenstates to a local perturbation. We characterized the effect of a local perturbation by a dimensionless parameter $\mathcal{G}$, defined via the ratio of a matrix element of local operator to the level spacing [Eq.~(\ref{eq:g})]. This parameter is reminiscent of the Thouless conductance in the single-particle localization problem, which describes the response of energy levels to changing the boundary conditions. Similar to the single-particle case, we argued that the transition can be identified by the relation $\mathcal{G}(L)\approx \mathcal{G}_c \sim 1$. However, it should be noted that, unlike the Thouless conductance, the parameter $\mathcal{G}$ is not directly related to the physical conductance of a many-body system. 

We have verified our approach by studying the MBL and delocalization transition in the random-field XXZ model. We obtained an estimate for the critical disorder strength at the transition which agrees very well with previous numerical studies based on entanglement and energy level statistics. Further, we mapped out the mobility edge in this model, demonstrating that the method has a good energy resolution, while at the same time requiring little beyond exact diagonalization of the Hamiltonian. 

In addition to being an efficient tool to detect the MBL transition, our criterion gives insights into its microscopic details. In particular, our criterion implies that at the MBL transition, both transport of conserved quantities (such as spin polarization), as well as spreading of entanglement,  are logarithmically slow. Such dynamics at the MBL transition was predicted by recent phenomenological RG studies of the MBL transition~\cite{AltmanRG14,Potter15}. Using exact diagonalization and TEBD calculations, we have computed the spreading of entanglement and transport in large systems, confirming the logarithmic in time behaviour for both quantities.

Finally, we mention some further directions opened by this study. First, we expect that this method will be useful for studying MBL and ergodicity breaking in other contexts, for example in the translationally-invariant models which were conjectured to break ergodicity~\cite{Huveneers13,Muller, DeRoeck,*DeRoeckArxiv, Garrahan1, Yao14, Papic, Garrahan2}. Second, our results provide a natural starting point for developing a microscopic renormalization group (RG) procedure for the MBL transition. In spirit, the microscopic parameter ${\cal G}$ is similar to the phenomenological parameter  (ratio of the ``entangling rate'' to the level spacing), introduced in a recent RG study~\cite{AltmanRG14}, while having the advantage of being numerically measurable. Hence, one could potentially use the ``physical'' distribution of $\cal G$, obtained from exact diagonalization, as an input parameter, and make use of the RG procedure to reach larger length scales. This would complement the existing phenomenological RG studies and is left for future work.  Another interesting direction would be to further explore the consequences of the LIOMs in the MBL phase. As we discussed above, our results hint at an underlying ultrametric structure, which could allow one to introduce a random matrix ensemble from which the properties of the MBL phase and the delocalization transition could be analytically computed.

\emph{Acknowledgments}.--- We acknowledge helpful discussions with Sid Parameswaran, Andrew Potter, Antonello Scardicchio, Romain Vasseur, and especially with Ehud Altman and David Huse. We would like to thank Miles Stoudenmire for the assistance with ITensor library. Research at Perimeter Institute is supported by the Government of Canada through Industry Canada and by the Province of Ontario through the Ministry of Economic Development \& Innovation.
This research was  supported by Gordon and Betty Moore Foundation's EPiQS Initiative through Grant GBMF4307 (M.S.), Sloan Foundation, NSERC and Early Researcher Award of Ontario (D.A.). This work made use of the facilities of N8 HPC provided and funded by the N8 consortium and EPSRC (Grant No.EP/K000225/1).

\end{document}